Title: Rotation Numbers, Boundary Forces and Gap Labelling
Authors: J. Kellendonk, I.P. Zois
Comments: 10 pages, final version accepted for publication
Report-no:
Subj-class: mathematical physics
Journal-ref: J. Phys. A: Math. Gen. 38 (2005) 3937-3946 

We review the Johnson-Moser rotation number and the $K_0$-theoretical
gap labelling of Bellissard for one-dimensional Schr\"odinger 
operators.
We compare them with two further gap-labels, one being related to the
motion of Dirichlet eigenvalues, the other being a $K_1$-theoretical
gap label. We argue that the latter provides a natural generalisation
of the Johnson-Moser rotation number to higher dimensions.

\documentclass[12pt,a4paper]{article}
\usepackage{amssymb,amsmath,latexsym}
\title{Rotation Numbers, Boundary Forces and  Gap Labelling}

\author{Johannes Kellendonk${}^1$, Ioannis Zois${}^2$ 
\\ 
\\
$^1$ {\small Institute Girard Desargues, Universit\'e Claude
 Bernard Lyon 1, F-69622
 Villeurbanne} \\ 
{\small e-mail: kellendonk@igd.univ-lyon1.fr} \\
{\small ${}^2$ School of Mathematics, Cardif University, PO Box 926, 
Cardiff
 CF24 4YH, UK} \\ 
{\small e-mail: zoisip@cf.ac.uk}}

\date{January 31, 2005}

\newtheorem{thm}{Theorem}
\newtheorem{defn}{Definition}

\newtheorem{lem}{Lemma}
\newtheorem{cor}{Corollary}
\newtheorem{rem}{Remark}

\newcommand{\RR}{\mathbb R}

\newcommand{\CC}{\mathbb C}
\newcommand{\Z}{\mathbb Z}

\newcommand{\Tr}{\mbox{\rm Tr}}  
\newcommand{\TV}{{\mathcal T}}
\newcommand{\TVh}{\hat{{\mathcal T}}}

\newcommand{\bew}{{\bf Proof:}}
\newcommand{\eb}{\hfill $\Box$}

\newcommand{\bs}{\bigskip}

\newcommand{\talpha}{{\tilde\varphi}}

\newcommand{\CA}{$C^*$-algebra}
\newcommand{\SO}{Schr\"odinger operator}

\newcommand{\hull}{\Omega}

\newcommand{\rD}{right Dirichlet value}
\newcommand{\lD}{left Dirichlet value}

\newcommand{\IDS}{{\rm IDS}}
\newcommand{\PP}{{\bf P}}

\newcommand{\rot}{{\rm rot}}
\newcommand{\Uu}{\mathcal U}
\begin{document}

\maketitle

\begin{abstract}
We review the Johnson-Moser rotation number and the $K_0$-theoretical
gap labelling of Bellissard for one-dimensional Schr\"odinger 
operators.
We compare them with two further gap-labels, one being related to the
motion of Dirichlet eigenvalues, the other being a $K_1$-theoretical
gap label. We argue that the latter provides a natural generalisation
of the Johnson-Moser rotation number to higher dimensions.
\end{abstract}

\section{Introduction}

It is an interesting and well known observation that the boundary of a
domain plays a prominent role both in mathematics and in physics.
A case that comes immediately into mind is the theory of
differential equations where the boundary conditions 
determine quite a lot of the whole solution.
In a purely topological context 
the boundary may even determine
the behaviour of the system in the bulk completely. 
A case like this was studied in
\cite{KS1,KS2} where a correspondance
between bulk and boundary topological invariants for certain
physical systems arising in solid state physics was found.
This 
was mathematically based on 
$K$-theoretic and cyclic cohomological
 properties of the
Wiener-Hopf extension of the $C^{*}$-algebra of observables. In most
applications we have in mind, this $C^{*}$-algebra is
obtained by considering the Schr\"odinger operator and its
translates describing the $1$-particle approximation of the solid.\\
In this
article we consider a simple example, a Schr\"odinger operator on the
real line, where such a correspondance can be established more
directly with the help of the Sturm-Liouville theorem.
The $K_0$-theory gap labels (below referred to also as even $K$-gap 
labels)
introduced by Bellissard et al.\
\cite{BLT,Be92} are bulk invariants. 
It is known that these are equal to the Johnson-Moser
rotation numbers \cite{jm} 
the existing proof being essentially a corollary of the
Sturm-Liouville theorem by which they are identified
with the integrated density of states on the gaps. 
In the first part of the paper (Sections~2,3)
we provide a direct identification
 of the Johnson-Moser rotation number (for energies in gaps)
with a boundary invariant, here called the
Dirichlet rotation number. This boundary invariant has
a physical interpretation, namely as boundary force per unit energy. 
Moreover, it can be interpreted as a $K_1$-theory gap label (or odd
$K$-gap label). 

In the second part (Sections~4,5) we indicate how 
the equality between the $K_0$ and the
$K_1$-theory gap labels also follows from 
the above-mentioned noncommutative topology of
the Wiener Hopf extension. The advantage of this approach is that,
unlike the definition of the geometrical rotation numbers and the
Sturm-Liouville theorem, it
is not restricted in dimension. We tend to think of the $K_1$-theory
gap label, which is naturally defined in any dimension,
as the operator algebraic formulation of the 
Johnson-Moser rotation number.

Whereas the first part is based on a single
operator, although its translates play a fundamental role, we consider
in the second part covariant families of operators indexed by the hull
of the potential. This is the right framework for the use of ergodic
theorems and noncommutative topology.
The last section is mainly based on \cite{Ke04b} and therefore
held briefly.

\section{Preliminaries}

In this article we consider as in \cite{Johnson86} a one-dimensional
\SO\ $H=-\partial^2 + V$ with (real) bounded potential which we assume 
(stricter as in \cite{Johnson86}) to be bounded differentiable. 
We also consider its translates $H_\xi := -\partial^2 + V_\xi$,
$V_\xi(x) = V(x+\xi)$, and lateron its hull. 
The differential equation $H\Psi = E\Psi$ 
for complex valued functions $\Psi$ over $\RR$
has for all $E$ two linear independent solutions but not all $E$
belong to the spectrum $\sigma(H)$ of $H$ as an operator acting on
$L^2(\RR)$. In this situation 
the following property of solutions 
holds \cite{CL55}.
\begin{thm}\label{thm3}
If $E\notin\sigma(H)$ there exist two real solutions $\Psi_+$ and 
$\Psi_-$ of
$(H-E)\Psi=0$, $\Psi_+$ vanishing at $\infty$ and $\Psi_-$ vanishing
at $-\infty$. 
These solutions are linear
independent and unique up to multiplication by a factor. 
\end{thm}

We mention as an aside that Johnson proves even
exponential dichotomy for such energies 
\cite{Johnson86}.  
Clearly $\sigma(H_\xi)=\sigma(H)$ for all $\xi$.

We consider also the action of $H_\xi$ 
on $L^2(\RR^{\leq 0})$ 
with Dirichlet boundary conditions at the boundary. 
If we need to emphazise this we will also write
$\hat H_\xi$ for the half-sided operator. The spectrum is then no 
longer the
same. Whereas the essential part of the spectrum of $\hat H_\xi$ is 
contained in that of $H_\xi$ \cite{Johnson86} 
the half sided operator may have isolated
eigenvalues in the gaps in $\sigma(H_\xi)$. 
Here a gap is a connected component of the complement of the spectrum,
hence in particular an open set.
$E$ is an eigenvalue of $\hat H_\xi$
if $(\hat H_\xi - E)\Psi = \Psi$ for $\Psi\in L^2(\RR^{\leq 0})$ which 
for
$E$ in a gap of $\sigma(H_\xi)$ amounts to saying that the solution
$\Psi_-$ of $(H_\xi-E)\Psi_-=0$ from Theorem~\ref{thm3} satisfies 
in addition $\Psi_-(0)=0$. 
\begin{defn}
We call $E\in\RR$ a \rD\ of $H_\xi$ if it is an eigenvalue of  
$\hat H_\xi$.
\end{defn}

We recall the important Sturm-Liouville theorem:
 \begin{thm}\label{thm-SL}
 Consider $H:=-\partial^2 + V$ with (real) bounded continuous
 potential acting on $L^2([a,b])$ with Dirichlet boundary conditions.
 The spectrum is discrete and bounded from below. 
A real eigenfunction to the $n$th eigenvalue (counted from below)
 has exactly $n-1$ zeroes in the interior $(a,b)$ of $[a,b]$.
 \end{thm}

\section{Rotation numbers}\label{sec-3}

The winding number of a continuous function $f:\RR/\Z\to \RR/\Z$
is intuitively speaking the number of times its graph
wraps around the circle $\RR/\Z$. This is counted relative to the
orientations induced by the order on $\RR$.
Let $\Lambda=\{\Lambda_n\}_n$ be 
an increasing chain of compact intervals 
$\Lambda_n=[a_n,b_n]\subset\Lambda_{n+1}\subset\RR$ 
whose union covers $\RR$.
The quantity 
$$\Lambda(f) := \lim_{n\to \infty} \frac{1}{b_n-a_n}\int_{a_n}^{b_n}
f(x) dx $$
is called the $\Lambda$-mean of the function $f:\RR\to\RR$, 
existence of the limit assumed. 
Now let $f:\RR\to \RR/\Z$ be continuous
and choose a continuous extension $\tilde f:\RR\to \RR$. 
To define the rotation number of $f$ we consider the expression
$$\rot_{\Lambda}(f) = 
\lim_{n\to \infty} \frac{\tilde f(b_n)-\tilde f(a_n)}{b_n-a_n} $$
which becomes the winding number of $f$ if $f$ is periodic of period 
$1$. 
The limit does not exist in general
but if it does it is independent of the extension $\tilde f$.
If $f$ is piecewise differentiable then
$\rot_{\Lambda}(f) = \Lambda(f')$. 
Moreover, if 
$U:\RR \to \CC$ is a nowhere vanishing continuous piecewise
differentiable function then we can 
consider the rotation number of its argument function which becomes
\begin{equation}\label{eq-rot} 
\rot_{\Lambda}(\frac{\arg(U)}{2\pi}) = 
\lim_{n\to \infty} \frac{1}{2\pi i(b_n-a_n)}\int_{a_n}^{b_n} 
\frac{\overline{U}}{|U|} 
\left(\frac{U}{|U|}\right)'dx 
\end{equation}

\subsection{The Johnson-Moser rotation number}\label{sec-3.1}

Johnson and Moser in \cite{jm} have defined rotation numbers for
the \SO\ $H=-\partial^2 + V$ on the real line where $V$ is a real 
almost periodic potential. 
They are defined as follows: Let $\Psi (x)$ be the nonzero real 
solution
of $(H-E)\Psi =0$ which vanishes at $-\infty$, 
then $\Psi'+i\Psi:\RR\to \CC$ is nowhere vanishing and
\begin{equation}\label{def-rot}
\alpha_\Lambda(H,E):=2\,\rot_{\Lambda}(\frac{\arg(\Psi'+i\Psi)}{2\pi}) 
.
\end{equation}
(Our normalisation differs from that in \cite{jm} for later
convenience.)
For the class of potentials considered here the limit is indeed defined
and even independent on the choice of $\Lambda$, we will come back to
that in Section~\ref{sec-hulls}. 
 
Note that $\alpha_\Lambda(H,E)$ has the following 
interpretations. If $N(a,b;E)$ denotes the
number of zeroes of the above solution $\Psi$ in $[a,b]$ 
then $\alpha_\Lambda(H,E)$ is the $\Lambda$-mean of the density of 
zeroes of
$\Psi$, namely one has
$$\alpha_\Lambda(H,E)=\lim_{n\to\infty}
\frac{N(a_n,b_n;E)}{b_n-a_n} .$$
The integrated density of states of $H$ at $E$ is
\begin{equation}\label{IDS}
\IDS_\Lambda (H,E) = \lim_{n\to\infty}
\frac{1}{|\Lambda_n|}\Tr(P_E(H_{\Lambda_n})) 
\end{equation}
provided the limit exists.
Here $|\Lambda_n|=b_n-a_n$ is the volume of $\Lambda_n$, 
$H_{\Lambda_n}$ the restriction of $H$ to $\Lambda_n$
with Dirichlet boundary conditions and, for self-adjoint $A$,
$P_E(A)$ is the spectral projection 
onto the spectral subspace of spectral values smaller or
equal to $E$. It will be important that $P(A)$ is a continuous
function of $A$ if $E$ is not in the spectrum of $A$.
Since $\Tr(P_E(H_{\Lambda_n}))$ is the number of eigenfunctions of
$H_{\Lambda_n}$ to eigenvalue smaller or equal $E$
Theorem~\ref{thm-SL} implies 
\begin{cor}
$ \alpha_\Lambda(H,E)= \IDS_\Lambda(H,E)$.
\end{cor}
In particular, like the integrated density of states
$\alpha_\Lambda(H,E)$ is monotonically increasing in $E$ and constant
on the gaps of the spectrum of $H$. It is moreover the same for all 
$H_\xi$.

\subsection{The Dirichlet rotation number}\label{sec-3.2}

We now consider the
continuous 1-parameter family of operators $\{H_{\xi}\}_{\xi}$ with
$\xi\in \RR$
 and $H_{\xi}=-\partial ^{2}+V_{\xi}$, where $V_{\xi}(x)=V(x+\xi)$. We 
shall
prove that the Johnson-Moser rotation number is a rotation number which 
is
defined by \rD s as a function of $\xi$.

We choose a gap $\Delta$ in $\sigma(H_\xi)=\sigma(H)$ for this section
and define the set of \rD s in $\Delta$ 
$$D_\xi(\Delta):=\{\mu\in\Delta | \exists \Psi:\, (H_{\xi}-\mu)\Psi=0
\mbox{ and }\Psi(0)=\Psi(-\infty)=0\}\;.$$
Thus
with respect to this choice of gap we can define 
$$S(\mu):=\{\eta |\mu\in D_\eta(\Delta)\}.$$
Suppose $\mu\in D_\xi(\Delta)$ for some $\xi$ 
(in particular, $D_\xi(\Delta)\neq\emptyset$).
Then there exists  
a non-zero solution $(H_\xi-\mu)\Psi=0$ satisfying 
$\Psi(0)=\Psi(-\infty)=0$. Let
$$Z(\mu,\xi):=\{x | \Psi(x-\xi )=0\}.$$
This set depends actually only on $\mu$, since $\Psi$ is unique up to
a multiplicative factor and we have:
\begin{lem}\label{lem-2} Let $\xi\in\RR$ such that
$D_\xi(\Delta)\neq\emptyset$ and
$\mu\in D_\xi(\Delta)$. Then $S(\mu)=Z(\mu,\xi)$.
\end{lem}
\bew\
Let $\Psi$ be a non-zero solution $(H_\xi-\mu)\Psi=0$ satisfying 
$\Psi(0)=\Psi(-\infty)=0$ and
define $\Psi_\eta(x)=\Psi(x+(\eta-\xi))$. Then
$(H_\eta-\mu)\Psi_\eta=0$ and $\Psi_\eta(-\infty)=0$ for all $\eta$.
Hence 
$Z(\mu,\xi)=\{\eta|\Psi(\eta-\xi)=0\}=
\{\eta|\Psi_\eta(0)=0\}\subset S(\mu)$. 

For the opposite inclusion if $\mu\in D_\eta(\Delta)$, then there 
exists
$\Phi$  such that $(H_{\eta}-\mu)\Phi =0$ with 
$\Phi (0)=\Phi(-\infty)=0$. Define $\Phi_\xi(x+(\eta-\xi))=\Phi(x)$.
Then $(H_{\xi}-\mu)\Phi_\xi =0$ with $\Phi_\xi(-\infty)=0$.
By Theorem~\ref{thm3}, 
$\Psi =\lambda\Phi_{\xi}$ for some $\lambda\in\CC^{*}$, which
implies $\Psi(\eta-\xi)=\lambda\Phi(0)=0$ and hence $\eta\in 
Z(\mu,\xi)$, thus
$S(\mu)\subseteq Z(\mu,\xi)$.\eb\bs

Let $\xi\in S(\mu)$, $\mu\in\Delta$. 
Since the spectrum of $\hat H_\xi$ in the gap 
$\Delta$ consists of isolated eigenvalues which are non-degenerate by 
Theorem~\ref{thm3} we can use  
perturbation theory to find a neighbourhood
$(\xi-\epsilon,\xi+\epsilon)$ and a differentiable function
$\xi\mapsto\mu(\xi)$ on this neighbourhood which is uniquely defined
by the property that $\mu(\xi)\in D_\xi(\Delta)$. 
In fact, level-crossing of \rD s cannot occur in gaps, since it would 
lead to 
degeneracies.  
As in \cite{Ke04} we see that its first derivative is strictly 
negative:
$$\frac{d\mu(\xi)}{d\xi}=\int_{-\infty}^{0}dx|\Psi_{\xi}(x)|^{2}V_{\xi}'=-
|\Psi_{\xi}'(0)|^{2}<0.$$ 
Here $\Psi_{\xi}$ is a normalised eigenfunction of $\hat H_\xi$.
Thus around each value $\xi$ for which we find a
\rD\ in $\Delta$ we have locally defined curves $\mu(\xi)$ which are
strictly monotonically decreasing and non-intersecting. 
Since $\hat H_\xi$ is norm-continuous in $\xi$ in the generalised 
sense, 
its spectrum $\sigma(\hat H_\xi)$ is lower semi-continuous \cite{Kato}
in $\xi$ so that the curves $\mu(\xi)$ can
be continued until they reach the boundary of $\Delta$ or their limit
at $+\infty$ or $-\infty$, if it exists. 

Let $K$ be the circle of
complex numbers of modulus $1$. We define
the function $\tilde\mu:\RR\to K$ by
$$\tilde\mu(\xi)= 
\exp \left( 2\pi i\sum_{\mu \in D_\xi} \frac{\mu-E_0}{|\Delta|}\right) 
$$
where $E_0=\inf\Delta$ and $|\Delta|$ is the width of $\Delta$. 
Then $\tilde\mu$ is a continuous function which is differentiable at
all points where none of the curves $\mu(\xi)$ touches the boundary.
\begin{defn}
The Dirichlet rotation number is 
$$\beta_\Lambda(H,\Delta):=-\rot_\Lambda(\frac{\arg\tilde\mu}{2\pi})\;
.$$
\end{defn}

\begin{lem}\label{lem-3} If, for some $\mu\in\Delta$, $|S(\mu)|>1$ then  
$\Delta$ contains at most one \rD\ of $H_\xi$.
\end{lem}
\bew\ We first remark that the 
same discussion can be performed for the \lD s of $H_\xi$, namely 
values $E$ for which exist $\Psi$ solving $(H_\xi-E)\Psi=0$ with
$\Psi (0)=\Psi(+\infty)=0$. 
These similarily define locally curves $\mu^{*}(\xi)$
whose first derivative are now strictly positive. They can't
intersect with any of the curves $\mu(\xi)$,
because a \rD\ which is at the same time a \lD\ must be a
true eigenvalue of $H$. 
Let $S^*(\mu)$ and $Z^*(\mu)$ be defined as $S(\mu)$ and $Z(\mu)$
but for \lD s. 
We claim that between two points of $S(\mu)$ lies one point of 
$S^*(\mu)$.
This then implies the lemma, because if $D_\xi$ contained two points
an elementary geometric argument shows that
the curves defined by \rD s through these points necessarily have
to intersect a curve defined by \lD s.
To prove our claim we consider the analogous statement for
$Z(\mu)$ and $Z^*(\mu)$ and let 
$\Psi_\pm$ be a real solution of $(H_0-\mu)\Psi=0$
with $\Psi_\pm(\pm\infty)=0$. Since $\mu$ is not an eigenvalue the
Wronskian $[\Psi_+,\Psi_-]$ which is always constant does not vanish. 
Furthermore, if $\Psi_+(x)=0$ then
$\Psi_-(x)=-[\Psi_+,\Psi_-]/\Psi'_+(x)$. This expression changes sign
between two consecutive zeroes of $\Psi_+$ and hence $\Psi_-$ must
have a zero in between.\eb\bs

\begin{rem} {\rm Under the hypothesis of the lemma the sum in the
definition of $\tilde\mu$ contains at most one element.
We believe that the result of the lemma is true under
all circumstances.}
\end{rem}

\begin{thm} 
$\alpha_\Lambda (H,E) = \beta_\Lambda (H,\Delta)$.
\end{thm}
\bew\ By Lemma~\ref{lem-2} $\alpha_\Lambda (H,\mu)$ is the 
$\Lambda$-mean of the
density of  $S(\mu)$.  
Suppose the hypothesis of the Lemma~\ref{lem-3} holds. 
Then $S(\mu)$ can be identified with the set of
intersection points between the constant curve 
$\xi\mapsto \exp 2\pi i\frac{E-E_0}{|\Delta|}$ and
$\tilde\mu(\xi)$. Since $\mu'(\xi)<0$ 
the $\Lambda$-mean of the density of these intersection points is minus
the rotation number of $\frac{\arg\tilde\mu}{2\pi}$.

Now suppose that $S(\mu)$ contains at most one element.
Then $\alpha_\Lambda(H,\mu)=0$. On the other hand, there can only be 
finitely
many curves defined by \rD s. Since they intersect the constant curve 
$\xi\mapsto \exp 2\pi i\frac{\mu-E_0}{|\Delta|}$ only once, 
$\beta_\Lambda(H,\Delta)$ must be $0$.
\eb\\

\noindent {\bf Remark 1} 
An even nicer geometric picture arrises if we take into account also
the \lD s of $H_\xi$ for the definition of $\tilde\mu$.
For this purpose redefine  $\tilde\mu:\RR\to K$ by
$$\tilde\mu(\xi)= 
\exp \pi i \left(\sum_{\mu\in D_\xi} \frac{\mu-E_0}{|\Delta|} 
- \sum_{\mu\in D_\xi^*} \frac{\mu-E_0}{|\Delta|} \right) $$
where $D_\xi(\Delta)^*$ is the set of \lD s of $H_\xi$ in $\Delta$.
Then $\tilde\mu$ is as well a continuous piecewise differentiable 
function and
$\rot_\Lambda(\frac{\arg\tilde\mu}{2\pi})$ is the same number as
before except that it yields the $\Lambda$-mean of the winding per
length of the Dirichlet values around a circle which is obtained from
two copies of $\Delta$ by identification of their boundary
points. For periodic systems,
this circle can be identified with the homology cycle
corresponding to a gap in the complex spectral curve of $H$ \cite{M}
and so $\beta_\Lambda(H,\Delta)$ 
is the winding number of the Dirichlet values around it. 
This is similar to Hatsugai's interpretation of the edge Hall 
conductivity
as a winding number (see \cite{Hat93}). There the role of the
parameter $\xi$ is played by the magnetic flux.\\

\subsection{Odd $K$-gap labels and Dirichlet rotation 
numbers}\label{sec-3.3}

We define another type of gap label which is formulated using
operator traces and derivations instead of curves on topological
spaces. It has its origin in an odd pairing between
$K$-theory and cyclic cohomology.

We fix a gap $\Delta$ in the spectrum of $H$ of length $|\Delta|$ and 
set 
$E_{0}=\inf (\Delta)$. Let $P_{\Delta}=P_{\Delta}(\hat H_\xi)$ be the
spectral projection of $\hat H_\xi$ onto the energy interval $\Delta$.
Then
\begin{equation}\label{eq-U}
\Uu_\xi :=P_{\Delta}e^{2i\pi \frac{\hat H_{\xi}-E_{0}}{|\Delta
    |}}+1-P_{\Delta}
\end{equation}
acts essentially as the unitary of time evolution by time 
$\frac{1}{|\Delta|}$
on the eigenfunctions of $\hat H_\xi$ in $\Delta$. These eigenfunctions 
are all
localised near the edge and therefore is the following expression a 
boundary 
quantity. 
\begin{defn} The odd $K$-gap label is
$$\Pi_{\Lambda}(H,\Delta)=
-\lim_{n\to\infty}\frac{1}{2i\pi|b_n-a_n|}\int_{a_n}^{b_n}
\Tr[(\Uu^{*}_{\xi}-1)\partial _{\xi}\Uu_\xi]d\xi$$
Where $\Tr$ is the standard operator trace on $L^2(\RR)$.
\end{defn}
\begin{thm}
$\Pi_\Lambda(H,\Delta) = \beta_\Lambda(H,\Delta)$.
\end{thm}
\bew\
Note that the rank of $P_\Delta$ is equal to $|D_\xi(\Delta)|$, the 
number of
elements in $D_\xi(\Delta)$. Let us first suppose that this is either
$1$ or $0$ which would be implied under the conditions of 
Lemma~\ref{lem-3}.
Since $\Uu^{*}_{\xi}-1=P_{\Delta} (e^{2i\pi \frac{\hat 
H_{\xi}-E_{0}}{|\Delta
    |}} -1 )$ we can express the trace using the normalised
eigenfunctions $\Psi_\xi$ of $\hat H_\xi$ to
$\mu(\xi)$, provided $|D_\xi(\Delta)|=1$,
\begin{equation}\label{eq-tr}
\Tr[(\Uu^{*}_{\xi}-1)\partial _{\xi}\Uu_\xi] 
_{\xi}\Uu_\xi)|\Psi_\xi\rangle\\
= \langle\Psi_\xi| \Uu^{*}_{\xi}-1
|\Psi_\xi\rangle \langle\Psi_\xi|
\partial _{\xi}\Uu_\xi|\Psi_\xi\rangle.
\end{equation}
Substituting
$$\langle\Psi_\xi|\partial_{\xi}\Uu_\xi|\Psi_\xi\rangle =
\partial_\xi \langle\Psi_\xi|\Uu_\xi|\Psi_\xi\rangle
=\partial_\xi e^{2i\pi \frac{\mu(\xi)-E_{0}}{|\Delta |}} $$ 
in the previous expression 
we arrive at
$$ \Tr[(\Uu^{*}_{\xi}-1)\partial_{\xi}\Uu_\xi] = 
(e^{-2i\pi \frac{\mu(\xi)-E_{0}}{|\Delta |}}-1)\partial_\xi 
e^{2i\pi \frac{\mu(\xi)-E_{0}}{|\Delta |}}.$$
Since $\Uu_\xi^*-1=0$ if $D_\xi(\Delta)=\emptyset$ we have
\begin{equation}\label{eq-bf} \Pi_\Lambda(H,\Delta) 
= -\lim_{n\to\infty}\frac{1}{2i\pi|b_n-a_n|}\int_{a_n}^{b_n}
(\overline{\tilde\mu(\xi)}-1) \tilde\mu'(\xi)  d\xi =
-\frac{1}{2i\pi}\Lambda(\overline{\tilde\mu}\tilde\mu')
\end{equation}
which is the expression for $\beta_\Lambda(H,\Delta)$.

If $|D_\xi|>1$ one has to replace the r.h.s.\ of
(\ref{eq-tr}) by a sum over eigenfunctions of $\hat H_\xi$ and the
calculation will be similar.
\eb

\subsection{Interpretation as boundary force per unit 
energy}\label{sec-3.4}

We assume for simplicity $|D_\xi|\leq 1$. Then 
we obtain from (\ref{eq-bf})
$$
\Pi_\Lambda(H,\Delta)=-  
 \lim_{n\to\infty}\frac{1}{|b_n-a_n|}\int_{a_n}^{b_n} 
\mu'(\xi)\frac{|D_\xi(\Delta)|}{|\Delta|} d\xi \;.$$
The r.h.s.\ is $\frac{1}{|\Delta|}$ times
the $\Lambda$-mean of the expectation value
of the gradient force w.r.t.\ the density matrix associated with the
egde states in the gap. Since translating $\hat H_\xi$ in $\xi$ is
unitarily equivalent to translating the position of the boundary,  
$\Pi$ can be seen as the force
per unit energy the edge states in the gap of the system exhibit on
the boundary \cite{Ke04b}.

\section{Hulls and ergodic theorems}\label{sec-hulls}

So far we have worked with a single potential and its translates. When
completed w.r.t.\ a natural metric topology this set of translates
yields a topological space, called the \emph{hull} of the potential. As it
has become apparent in recent years, many topological invariants of
the physical system depend mainly on the topology of this hull with
its $\RR$ action by translation of the potential. Besides, the use of
invariant ergodic probability measures on the hull allows to tackle
the problem of existence of the $\Lambda$-means in a probabilistic 
sense. 
It is therefore most
natural to interprete the results of the last section in the framework
of $\RR$-actions on hulls. This allows for a generalisation to higher
dimensional systems, to which the theorems of Section~2 do not extend.
  


Given a potential $V$ consider its hull
$$\hull = \overline{\{V_\xi|\xi\in\RR\}}\;,$$
which is the compactification of the set of
translates of $V$ in the sense of \cite{Johnson86,Be92}.
The action of $\RR$ by translation of the
potential extends to an action on $\hull$ by homeomorphisms which we
denote by $\omega\mapsto x\cdot\omega$. 
The elements of $\Omega$ may be identified with those 
real functions (potentials)
which may be obtained as limits of sequences of translates of $V$. 
We shall write $V_\omega$ for the potential corresponding to
$\omega\in\Omega$. If $\omega_0$ is the point of $\hull$ corresponding
to $V$ then $V_\xi=V_{-\xi\cdot\omega_0}$.
Also $V_{y\cdot\omega}(x)=V_{\omega}(x-y)$ and so
the family of Hamiltonians $H_\omega=-\partial^2+V_\omega$ is
covariant in the sense that $H_{x\cdot\omega}=U(x)H_\omega U^*(x)$
were $U(x)$ is the operator of translation by $x$.
The bulk spectrum is by definition the union of their spectra. 
 
The valididty of the following theorem, namely that $\hull$ carries
an $\RR$-invariant ergodic probability measure, can be verified
for many situations, see \cite{BHZ} for considerations relating it
to the Gibbs measure. 
\begin{thm}
Suppose that $(\hull,\RR)$ carries an invariant ergodic probability
measure $\PP$.
Let $\Delta$ be a gap in the bulk spectrum and $E\in\Delta$.
Then almost surely (w.r.t.\ this measure) the limits to define
$\alpha_\Lambda(H_\omega,E)$
and $\Pi_{\Lambda}(H_\omega,\Delta)$ 
exist and are independent of $\Lambda$ and
$\omega\in\hull$.  The almost sure value of
$\Pi_\Lambda$ is the $\PP$-average 
$$ \Pi(\Delta) = \frac{1}{2i\pi} \int_\hull d\PP(\omega)  
\Tr((\Uu^*_{\omega} -1) \delta^\perp \Uu_{\omega} )$$
where $(\delta^\perp f)(\omega)= 
\left.\frac{d f(t\cdot\omega)}{dt}\right|_{t=0}$ and $\Uu_\omega$ is
defined as in (\ref{eq-U}) with $\hat H_\omega$ in place of $\hat 
H_\xi$.
\end{thm}
\bew\
The crucial input is Birkhoff's ergodic theorem which allows to
replace 
$$\lim_{n\to\infty}\frac{1}{|\Lambda_n|}\int_{\Lambda_n}
F(x\cdot\omega) dx = \int_\Omega d\PP F(\omega)$$ for almost
all $\omega$ and any $F\in L^1(\Omega,\PP)$.
The corresponding construction for the rotation number $\alpha$ 
has been carried out in \cite{jm} for
almost periodic potentials
and for the more general set up in \cite{Johnson86,Be92}.
For $\Pi_\Lambda$ the relevant function is $F(\omega) =
\Tr((\Uu_\omega^*-1)\delta^\perp\Uu_\omega)$ which leads to the
expression of the almost sure value of $\Pi_\Lambda$.
\eb\bs



\section{$K$-theoretic interpretation}\label{sec-K}

The dynamical system $(\Omega,\RR)$ does not depend on the details
of $V$, but only on its spatial structure (or what may be called its
long range order). In fact, for systems whose atomic positions are
described by Delone sets there are methods to
construct the hull directly from this set, c.f.\ \cite{BHZ,FHK}.  
The detailled form of the potential is rather encoded in a 
continuous function $v:\hull\to\RR$ 
so that 
$V_\omega(x) = v(-x\cdot\omega)$ is the potential corresponding to 
$\omega$.
$C(\Omega)$ is thus the algebra of continuous potentials for a given
spatial structure.

If one combines this algebra with the Weyl-algebra of rapidly 
decreasing
functions of momentum operators 
one obtains the algebra of continuous observables which is
the $C^*$-crossed product $C(\hull)\rtimes_\varphi\RR$. It is the
$C^*$-closure of the convolution algebra of functions $f:\RR\to
C(\hull)$ with product 
$f_1f_2(x) = \int_\RR dy f_1(y)\varphi_{y}f_2(x-y)$ and involution 
$f^*(x) =
\varphi_{x}\overline{f(-x)}$, where
$\varphi_y(f)(\omega) = f(y\cdot\omega)$. It has a faithful
family of representations 
$\{\pi_\omega\}_{\omega\in\Omega}$ on $L^2(\RR)$ by integral operators,
$$\langle x|\pi_\omega(f)|y\rangle = f(y-x)(-x\cdot\omega).$$
It has the following important property.
For each continuous function $F:\RR\to\CC$ vanishing at $0$ and 
$\infty$
there exists an element $\tilde F\in C(\hull)\rtimes_\varphi\RR$ such 
that
$F(H_\omega) = \pi_\omega(\tilde F)$. 
Some of the topological properties of the family of Schr\"odinger
 operators $\{H_\omega\}_{\omega\in\hull}$ are therefore captured by
 the topology of the \CA.    
The invariant measure $\PP$ over $\Omega$ gives rise to a trace
$\TV:C(\Omega)\rtimes_\varphi\RR \to \CC$,
$\TV(f)=\int_\Omega d\PP f(0)$.
\begin{thm} [\cite{Be92}] Let $E$ be in a gap of the bulk spectrum
of  $\{H_\omega\}_{\omega\in\hull}$ so that in particular 
there exists a projection 
$\tilde P_E\in C(\hull)\rtimes_\varphi\RR$ such that  
$\pi_\omega(\tilde P_E) = P_E(H_\omega)$ is the projection onto the
spectral subspace of $H_\omega$ to energies below the gap.  
Suppose that the potential which gave rise to the hull $\Omega$ is 
smooth.
Then the almost sure value of $\IDS_\Lambda(H,E)$ is 
$\IDS(E):=\TV(\tilde P_E)$.
\end{thm}
We mention that this result is more subtle than just an application of
Birkhoff's theorem and interpretating the result in \CA ic terms
as it needs a Shubin type argument which holds for smooth potentials, 
namely
$$ \lim_{n\to \infty}
\frac{1}{|\Lambda_n|}\left(
\Tr(P_E(H_{\Lambda_n})-\Tr(\chi_{\Lambda_n}P_E(H))\right)
= 0.$$ 

The element $\tilde P_E$ is a projection. 
As any trace on a \CA, $\TV$ depends only on the homotopy class of
$\tilde P_E$ in the set of projections of 
$C(\Omega)\rtimes_\varphi\RR$.
The even $K$-group 
$K_0(C(\Omega\rtimes_\varphi\RR)$ is constructed from homotopy classes
of projections and 
the map on projections
$P\mapsto \TV(P)$ induces a functional on this
group, or stated differently, the elements of the $K_0$-group pair
with $\TV$. It is therefore reasonable to refer to $\TV(\tilde P_E)$ as an 
\emph{even}
$K$-gap label (or $K_0$-theory gap label) of the gap. 
This is the $K_0$-theoretical gap labelling
of \cite{BLT,Be92}.\\

There is a similar identification of 
the odd $K$-gap label as the result of a functional applied to the odd
$K$-group of a \CA. This \CA\ is the $C^*$-algebra of observables on
the half space near
$0$, the position of the boundary.
It turns out to be convenient to
consider also the cases in which the boundary is at $s\neq 0$.
We therefore consider the space $\Omega\times\RR$
with the product topology. 
This topological space, whose second component denotes the position of
the boundary, carries an
action of $\RR$ by translation of the potential and the
boundary (so that their relative position remains the same).
The relevant \CA\ is then the crossed product (constructed as above)
$C_0(\Omega\times\RR)\rtimes_{\talpha}\RR$ with
$\talpha_y(f)(\omega,s) = f(y\cdot\omega,s+y)$.
It has a family of representations 
$\{\pi_{\omega,s}\}_{\omega\in\Omega,s\in\RR}$ on 
$L^2(\RR)$ by integral operators,
$$\langle x|\pi_{\omega,s}(f)|y\rangle = f(y-x)(-x\cdot\omega,s-x).$$
It has the following important property:
for 
each continuous function $F:\RR\to\CC$ vanishing at $0$ and $\infty$
and such that $F(H_\omega)=0$ for all $\omega$,
there exists an element $\hat F\in
C_0(\Omega\times\RR)\rtimes_\talpha\RR$ such that
$F(H_{\omega,s}) = \pi_{\omega,s}(\hat F)$, where 
$H_{\omega,s}$ is the restriction of $H_\omega$ to $\RR^{\leq s}$ with
Dirichlet boundary conditions at $s$. Let $\Uu=\{\Uu_{\omega,s}\}$,
 \begin{equation}\label{eq-Us}
\Uu_{\omega,s} :=P_{\Delta}
e^{2i\pi \frac{H_{\omega,s}-E_{0}}{|\Delta |}}+1-P_{\Delta},
\end{equation}
similar to (\ref{eq-U}).
The product measure of $\PP$ with the Lebesgue measure is
an $\RR$-invariant measure on $\hull\times\RR$ and defines
a trace $\TVh(f) = \int_\hull\int_\RR d\PP ds f(0).$
\begin{thm}[\cite{Ke04b}] Let $\Delta$ be a gap in the bulk spectrum
of $\{H_\omega\}_{\omega\in\hull}$.
The almost sure value of $\Pi(\Delta)$ is 
$$ \Pi_\Lambda(H,\Delta)=\Pi(\Delta) := \frac{1}{2i\pi} 
\TVh(\widehat{\Uu^*-1} \delta^\perp \widehat{\Uu-1} ).$$
\end{thm}
The expression of the theorem depends only on the homotopy class of
$\widehat{\Uu-1} +1 $ in the set of unitaries of (the
unitization of)
$C_0(\Omega\times\RR)\rtimes_\talpha\RR$.
The odd $K$-group 
$K_1(C_0(\Omega\times\RR)\rtimes_\talpha\RR)$
is constructed from homotopy classes of unitaries and 
the map on unitaries
$U\mapsto \TVh((U^*-1)\delta^\perp U)$ induces a functional on this
group. It is therefore that we refer to $\frac{1}{2i\pi} 
\TVh(\widehat{\Uu^*-1} \delta^\perp \widehat{\Uu-1} )$
as an odd $K$-gap label of the gap.\\

The proof of the following theorem is based on the topology of 
the above \CA s.
\begin{thm}[\cite{Ke04b}] \label{thm5}
$\TV(\tilde P_E)= \frac{1}{2i\pi} 
\TVh(\widehat{\Uu^*-1} \delta^\perp \widehat{\Uu-1} )$. 
In other words, $\IDS(E)=\Pi(\Delta)$, $E\in\Delta$.
\end{thm}
\section{Conclusion and final remarks}

We have discussed four quantities which serve as gap-labels for
one-dimensional Schr\"odinger operators. They are all equal but their
definition relies on different concepts. The Johnson-Moser rotation
number $\alpha$ measures the mean oscillation of a single solution. The
Dirichlet rotation number $\beta$ counts the mean winding of the 
eigenvalues
of the halfsided operators around a circle compactification of the
gap. $\Pi$ and $\IDS$ are operator algebraic expressions with concrete
physical interpretations, the boundary force per energy and the
integrated density of states. Whereas the identities $\alpha=\beta=\Pi$
are rather elementary, their identity with $\IDS$ is based on a
fundamental theorem, the Sturm-Liouville theorem. We tend to think
therefore of $\Pi$ as the natural operator algebraic formulation of
the Johnson-Moser rotation number 
and of Theorem~\ref{thm5} as an operator analog of the Sturm-Liouville
theorem. The advantage is that $\Pi$, $\IDS$ and 
Theorem~\ref{thm5} generalise naturally to
higher dimensions \cite{Ke04b}. In fact, the expression for $\IDS$ is
the same as in (\ref{IDS}) if one uses F{\o}llner sequences 
$\{\Lambda_n\}_n$ for $\RR^d$.
The expression of $\Pi_\Lambda$ in $\RR^d$ 
requires a choice of a $d-1$-dimensional subspace, the boundary, and
so $\hat H_\xi$ is the restriction of the Schr\"odinger operator
$H_\xi=-\Sigma_j\partial_j^2 + V_\xi$, $V_\xi(x)=V(x+\xi e_d)$, to the
half space $\RR^{d-1}\times\RR^{\leq 0}$ with Dirichlet boundary 
conditions.  
Then 
$$ \Pi_\Lambda = -\lim_{n\to\infty}\frac{1}{|\Sigma_n|(b_n-a_n)}
\int_{a_n}^{b_n}
\Tr((\Uu^*_{\xi,\Sigma_n}-1)\partial_\xi \Uu_{\xi,\Sigma_n})d\xi\;, $$
$$\Uu_{\xi,\Sigma_n} = 
P_\Delta(\hat H_{\xi,\Sigma_n}) 
e^{2\pi i\frac{\hat H_{\xi,\Sigma_n}-E_0}{|\Delta|}} + 1 - 
P_\Delta(\hat H_{\xi,\Sigma_n})\; .$$
Here $\Sigma_n$ is a
F{\o}llner sequence for the boundary and 
$\hat H_{\xi,\Sigma_n}$ is the restriction
of $H_\xi$ to $\Sigma_n\times\RR^{\leq 0}$ with Dirichlet boundary 
conditions.
We do not know of a direct link between this expression and the
generalisation proposed by Johnson \cite{Johnson91} for odd-dimensional
systems.\\

\emph{Acknowledgements:}\\
The second author would like to thank EPSRC for financial support (contract number GR/R64995/01) and the 
University of Lyon I,  Institute Girard Desargues, for its hospitality.\\

\end{document}